# Layer Ensemble Averaging for Improving Memristor-Based Artificial Neural Network Performance

**Osama Yousuf**[1,2], Brian Hoskins[2], Karthick Ramu[2], Mitchell Fream[2], William A. Borders[2], Advait Madhavan[2], Matthew W. Daniels[2], Andrew Dienstfrey[3], Jabez J. McClelland[2], Martin Lueker-Boden[4], Gina C. Adam[1*]

[1] Department of Electrical and Computer Engineering, George Washington University, Washington, DC, 20052.

[2] National Institute of Standards and Technology, Gaithersburg, MD, 20899.

[3] National Institute of Standards and Technology, Boulder, CO, 80305.

[4] Western Digital Technologies, San Jose, CA, 95119.

* Corresponding author (email: ginaadam@gwu.edu)

**Abstract**

Artificial neural networks have advanced due to scaling dimensions, but conventional computing faces inefficiency due to the von Neumann bottleneck. In-memory computation architectures, like memristors, offer promise but face challenges due to hardware non-idealities. This work proposes and experimentally demonstrates layer ensemble averaging – a technique to map pre-trained neural network solutions from software to defective hardware crossbars of emerging memory devices and reliably attain near-software performance on inference. The approach is investigated using a custom 20,000-device hardware prototyping platform on a continual learning problem where a network must learn new tasks without catastrophically forgetting previously learned information. Results demonstrate that by trading off the number of devices required for layer mapping, layer ensemble averaging can reliably boost defective memristive network performance up to the software baseline. For the investigated problem, the average multi-task classification accuracy improves from 61 % to 72 % (< 1 % of software baseline) using the proposed approach.

**Introduction**

The increasing demand for large-scale neural network models has prompted a focused exploration of approaches to optimize model efficiency and accelerate computations. Quantized neural networks, which employ reduced-precision representations for model parameters and activations, have emerged as a promising avenue for achieving significant computational gains without compromising performance. Recent works push this field to extreme ends with demonstrations of effective 1-bit[1] and 1.58-bit[2] (ternary) quantization of the parameter space with minimal performance loss. As the community delves into extreme quantization, another frontier in enhancing neural network efficiency unfolds through the exploration of emerging memory-based hardware accelerators. This technology offers a synergistic solution that can complement efficiency gains achieved by quantized neural networks via in-memory computation on a physical chip of memory devices[3].

Memristors are two-terminal non-volatile memory devices that exhibit unique programmable resistive switching behavior. Their intrinsic characteristics enable co-location of computation and memory, mimicking aspects of synaptic functionality in biological systems[4,5]. By arranging memristors over a two-dimensional array of intersecting wires such that the devices are placed at intersection points, an architecture commonly referred to as a crossbar[6–8], underlying device physics can be exploited to implement parallelized vector-matrix multiplication – a critical operation in artificial neural networks – in the analog domain. Since these devices are programmable and non-volatile[9,10], they can be utilized to build dedicated hardware accelerators for deep neural networks. Diverse technologies including resistive random-access memory (ReRAM) and phase change memory (PCM) are being considered as promising candidates for using crossbar arrays to implement synaptic weights in neural networks[11–19]. Hardware accelerators based on these technologies can overcome von Neumann architecture limitations as they minimize data movement and energy consumption, both of which are key bottlenecks for large-scale neural network workloads. For these reasons, memristor-based neural network accelerators have the potential to transform capabilities of artificial intelligence and machine learning systems and thereby usher in a new neuromorphic era of intelligent edge computing. A comprehensive exploration of the interplay between quantized neural networks, dedicated hardware accelerators, and memristive technologies becomes imperative for advancing the capabilities of modern neural network workloads, with the overarching goal of unlocking unprecedented efficiency gains in real-world deep learning applications.

However, the study of memristor-based neural network accelerators faces several challenges. A purely experimental approach is unfeasible since commercial tape-outs have long timelines and significant design and fabrication costs[4,20,21]. Before hardware prototyping can be practically motivated, results from hardware-aware[22–24] simulations – simulations that take hardware characteristics into account – are needed to reliably predict the performance of these systems. The critical issue is that these devices can exhibit complex non-idealities such as cycle-to-cycle variability (inconsistency in performance from one switching cycle to the next), device-to-device variability (variation in behavior between individual devices), and even tuning failure (where the resistance state remains stuck at a certain level). These non-idealities manifest as deviations in the outputs of the underlying vector-matrix multiplications with the result that memristive neural networks may not achieve software-equivalent accuracies.

There have been several advancements in mitigating the impact of device non-idealities and in bridging the performance gap in inference accuracy of memristive networks compared to their software counterparts[6,25–30], with more recent works focusing on challenging continual learning settings[31,32]. Existing literature on the subject can be classified into two broad categories: the first focuses on circuit-level and device-level optimizations[6,25] such as alterations to the crossbar configuration and circuitry or the device material stack, and the second on network-level algorithmic optimizations[26–30] such as advanced programming or weight-to-device mapping schemes. Algorithmic investigations can be further divided into two sub-categories. The first utilizes information about hardware defects and attempts to train defect or hardware-aware solutions for the memristive hardware, while the second attempts to transfer a pure-software solution on the memristive hardware intelligently. Both approaches share the goal of maximizing the performance of the memristive network. The first achieves this by training redundancy into the solution, while the second achieves this by averaging out induced currents from multiple mapped instances of one or more solutions on the crossbar.

This work presents an algorithmic framework to utilize defective memristive crossbar devices for fully connected neural network inference. We present layer ensemble averaging – a novel weight-to-device mapping technique for achieving near-software hardware inference performance. Our algorithm mitigates the impact of device non-idealities at the level of individual neural network layers and thereby aligns the outputs of memristive hardware layers with software counterparts. Contrary to existing related literature[26,30] where neural network outputs are obtained by polling outputs of an ensemble of neural networks, not necessarily mapping the same solution, here ensemble outputs are polled at the level of each layer by mapping the same solution multiple times. This key difference makes our approach suitable for cases where the availability of software solutions is limited (perhaps due to high training costs or long training times) and one wishes to deploy a pre-trained model directly to memristive hardware for acceleration with minimal performance loss. It also allows for more accurate vector-matrix multiplication at the level of each layer and a straightforward way to investigate hardware outputs relative to software outputs of the vector-matrix multiplication operation for any given layer. Consequently, the proposed approach has the added benefit of being useful for domains beyond deep learning that require accurate multiply-and-accumulate or vector-matrix multiplication operations such as digital signal processing, image and video processing, scientific computing, and financial modeling.

In this work, we demonstrate that a full-precision neural network solution, pre-trained ex-situ in software for a continual learning, multi-task classification problem, quantized and written to a non-ideal memristive crossbar can attain near-software inference performance using the novel layer ensemble averaging framework. Choosing a continual learning problem provides a more rigorous evaluation of the robustness of the averaging scheme compared to simpler problems such as single-task classification. This is because continual learning effectively requires squeezing in more functionality into a network with a fixed capacity[33], thereby placing greater demands on the performance of layer ensemble averaging. We investigate the effectiveness of our approach in simulation with randomly created populations of defective crossbars as well as in experiment with an array of 20,000 non-ideal ReRAM devices using a novel mixed-signal hardware prototyping platform called Daffodil[34]. This prototyping system consists of a custom integrated chip with 20,000 memristive devices, a custom mixed-signal printed circuit board (PCB), a Zynq-based[a] field programmable gate array (FPGA) development board, and an accompanying software suite guided by principles of hardware-software co-design for hardware experiments as well as accurate system simulations.

---

[a] Certain commercial processes and software are identified in this article to foster understanding. Such identification does not imply recommendation or endorsement by the National Institute of Standards and Technology, nor does it imply that the processes and software identified are necessarily the best available for the purpose.

The novelty of this work is captured by the following key contributions. Firstly, we demonstrate the effectiveness of an averaging technique for mitigating the impact of device non-idealities for memristor-based artificial neural networks. Secondly, we showcase the Daffodil[34] mixed-signal prototyping system which has been developed for the memristive research community to standardize resistive neural network benchmarking. Using these algorithmic and experimental techniques, we successfully demonstrate the first fully experimental implementation of near-software inference (within 1 % of software baseline) on a continual learning or multi-task classification problem, to the best of our knowledge, using non-ideal ReRAM memristive crossbars. This work aims to not only propel our theoretical understanding of the synergies between memristors and neural networks in the context of memristive accelerators for deep learning, but also address practical needs of researchers seeking to circumvent common non-idealities of memristive neural network implementations.

**Methods**

**Neural Network Details**

The dataset used for this work is derived from the Yin-Yang dataset[35] containing 4-dimensional input features and 3 output classes: *Yin*, *Yang*, and *Dot*. Each sample in the dataset represents a point in a two-dimensional representation of the Yin-Yang symbol, and the task is to classify samples according to their position within the symbol. This dataset serves as a good proof-of-concept problem for early-stage hardware benchmarking efforts because it is small compared to alternative classic datasets, and it exhibits a clear gap between inference accuracies attainable by shallow networks or linear solvers compared to deep neural networks because of non-linear decision boundaries. The dataset is generated by randomly sampling values $a_i$ and $b_i$ in the feature domain $[0, 1]$ that satisfy a set of mathematical equations defining the Yin-Yang characteristic shape. To support learning for models without trainable biases, the values $(1 - a_i, 1 - b_i)$ are also included, leading to a representation $(a_i, b_i, 1 - a_i, 1 - b_i)$ in the 4-dimensional feature space. Using this dataset, a continual learning problem is obtained by converting it to a multi-task classification problem, shown in **Figure 1a**. It is derived by firstly broadening the feature domain to $[0, 2]$. Then, *Task 1* instances are generated using the same mathematical equations as the original dataset, and *Task 2* instances are generated by adding a constant offset (of 1) to instances produced by the same equations. In effect, this creates two individual Yin-Yang classification problems. This is motivated by choices in continual learning literature where problems of similar difficulty are studied for consistency[33,36,37]. For effective continual learning, a network must be trained on the tasks sequentially and must retain classification performance on *Task 1* as it learns to classify *Task 2*.

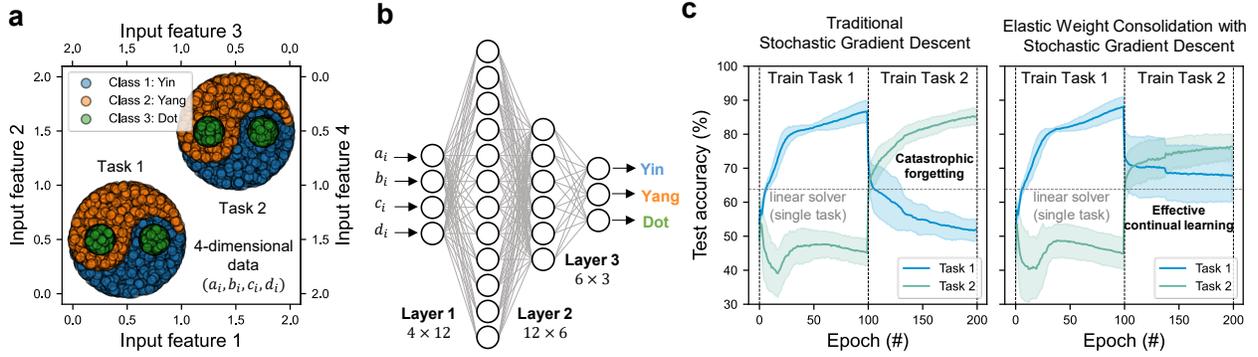

**Figure 1. Neural Network Details**. a) Visualization of the training set of the continual learning problem, which is a multi-task variant of the Yin-Yang dataset. b) Schematic of the 3-layer neural network architecture trained in software. c) Network convergence curves highlighting that the given problem and network exhibit the characteristic catastrophic forgetting problem when trained with stochastic gradient descent but demonstrate effective continual learning with elastic weight consolidation. The shaded region indicates standard deviation across 20 independent software networks starting at varying weight initializations.

A 3-layer fully connected perceptron network with no learnable biases is trained in software. The first layer has dimensions $4 \times 12$, the second layer has dimensions $12 \times 6$, and the third layer has dimensions $6 \times 3$, as shown in **Figure 1b**. For the input and hidden layers, a hyperbolic tangent activation function is used, and for the output layer, a softmax activation function is used. The final output classification is derived by performing an argmax operation over the 3 output neurons. The multi-task training process for 20 software networks starting at various weight initializations is presented in **Figure 1c**. For the first 100 epochs, the network is trained on instances from *Task 1*, and for the following 100 epochs, the same network is re-trained on instances from *Task 2*. The single-task linear solver test accuracy is included as a horizontal dotted line for reference to highlight where a deep neural network outperforms a linear classifier. When traditional stochastic gradient descent is used, the network exhibits the characteristic catastrophic forgetting problem, losing performance on the first task as it learns the second task, dropping below the performance of a linear solver. However, with elastic weight consolidation[33], the network can sufficiently maintain its performance on the previous task as it continues to learn, performing better than a linear solver on both tasks on average. Elastic weight consolidation is a fundamental technique for training deep neural networks in continual learning settings. At its core, elastic weight consolidation alters the training loss function by penalizing changes in weights that are important to previously seen tasks. As a result, it allows for effective continual learning in a supervised learning context.

We quantize these full-precision solutions trained in software using elastic weight consolidation to a ternary weight space using block reconstruction quantization (BRECQ)[38]. Then, we can directly utilize the quantized solutions for memristive hardware deployment and verification. For neural network results reported in this work, a single solution that performs better than the linear solver on both tasks is quantized and mapped to hardware layer ensembles for simplicity. Hyperparameters for training and quantization were optimized using Bayesian techniques with the Optuna framework[39]. Quantization-aware training schemes were investigated as well[40,41], but they failed to yield good continual learning results with elastic weight consolidation. We hypothesize this occurs because the loss landscape changes drastically when training deep networks in the ternary-weight domain, and elastic weight consolidation constraints restrict such highly quantized networks from effectively learning.

**Layer Ensemble Averaging**

We illustrate the principle of layer ensemble averaging in **Figure 2** for an example ternary weight matrix $\mathbf{W}$. Firstly, $\mathbf{W}$ is converted into two conductance matrices $\mathbf{G_{pos}}$ and $\mathbf{G_{neg}}$ such that $\mathbf{W} \propto (\mathbf{G_{pos}} - \mathbf{G_{neg}})$. A consequence of this differential encoding scheme is that software layer dimensions double when mapped to hardware i.e., each weight is represented by a pair of devices (one in $\mathbf{G_{pos}}$ and the other in $\mathbf{G_{neg}}$). For neural network demonstrations in this work, the ternary quantized weights $-1, 0, +1$ are represented by device pairs in conductance states $(G_{OFF}, G_{ON}), (G_{ON}, G_{ON})$ and $(G_{ON}, G_{OFF})$ respectively, and values of $G_{OFF}$ and $G_{ON}$ are chosen as 133 µS and and 233 µS based on current *vs.* voltage sweep measurements on the physical chip (presented later in **Figure 4**). For consistency, inputs are always applied on columns and outputs are always measured from the rows of the crossbar for this work. However, this is not a rigid requirement and can be altered as required. Each of the two conductance matrices is mapped to contiguous blocks of memristive devices multiple times as guided by our layer ensemble mapping algorithm (a layer ensemble refers to the complete hardware mapping of a layer). The primary parameter for layer ensemble mapping is $\beta$, which we define as the number of non-defective copies of rows. It directly controls how many candidates in a layer ensemble contribute to the current averaging process for a given output. A row is considered defective if it has one or more stuck devices, i.e. devices that cannot be tuned to a desired conductance state. For the given example, a layer ensemble mapping for $G_{pos}$ is shown with $\beta = 2$, indicating that the mapping consists of at least 2 non-defective copies of rows for each output. Note that the number of non-defective copies of a row ($\beta$) is different from the ensemble size ($\alpha$), which we define as the total number of copies of the matrix mapped on to the crossbar. In the example shown, the ensemble size $\alpha = 3$ indicates that the weight matrix (represented by conductance matrices $\mathbf{G_{pos}}$ and $\mathbf{G_{neg}}$) is mapped on to the crossbar three times, where $\mathbf{G_{pos}^i}$ represents the $i$-th mapping of $\mathbf{G_{pos}}$ (similarly for $\mathbf{G_{neg}}$). For the vector-matrix multiplication process, an input vector $\mathbf{X}$ can be converted to voltages by $\mathbf{V_{in}} = V_{read} \cdot \mathbf{X}$, where $V_{read}$ is the voltage used for reading operations on the crossbar. For this work, a low voltage of 0.3 V was used for reading to minimize unwanted disturbances to the memristor conductances. For each output, only currents from non-defective rows are considered for averaging. For the example in **Figure 2**, the layer ensemble output is thus given by,

$$\mathbf{I_{pos}} = [I_1 \quad I_2] = \left[\frac{i_1^2 + i_1^3}{\beta} \quad \frac{i_2^1 + i_2^2}{\beta}\right] \quad , \tag{1}$$

and the final vector-matrix multiplication output is given by,

$$\mathbf{XW} \approx \frac{\mathbf{V_{in}}}{V_{read}} \cdot \frac{(\mathbf{G_{pos}} - \mathbf{G_{neg}})}{[G_{norm} \cdot (G_{ON} - G_{OFF})]} = \frac{\mathbf{I_{pos}} - \mathbf{I_{neg}}}{[G_{norm} \cdot (G_{ON} - G_{OFF})] \cdot V_{read}} \quad , \tag{2}$$

where $\mathbf{I_{pos}}$ is the final output current vector from the ensemble mappings for $\mathbf{G_{pos}}$, $I_i$ is the averaged current for row $i$ produced by the layer ensemble, $i_i^j$ is the output current from the $j$-th mapping of $\mathbf{G_{pos}}$ for $I_i$, and $G_{norm}$ is a constant that scales the difference of the experimentally measured low conductance state $G_{OFF}$ from the high conductance state $G_{ON}$ averaged over devices present in the layer ensemble mapping. Averaging currents over defect-free $\beta$ rows mitigates the impact of device-to-device variability and skipping over the $\alpha - \beta$ defective rows for each output mitigates the impact of stuck and faulty devices, both of which can drastically reduce network performance if not accounted for as they cause hardware outputs to deviate from software. For hardware neural network results in this work, only the vector-matrix multiplication operations for each network layer are implemented on the physical chip. Other operations such as non-linear activation functions and current averaging are implemented in software using a separate host system (further detailed in the next section).

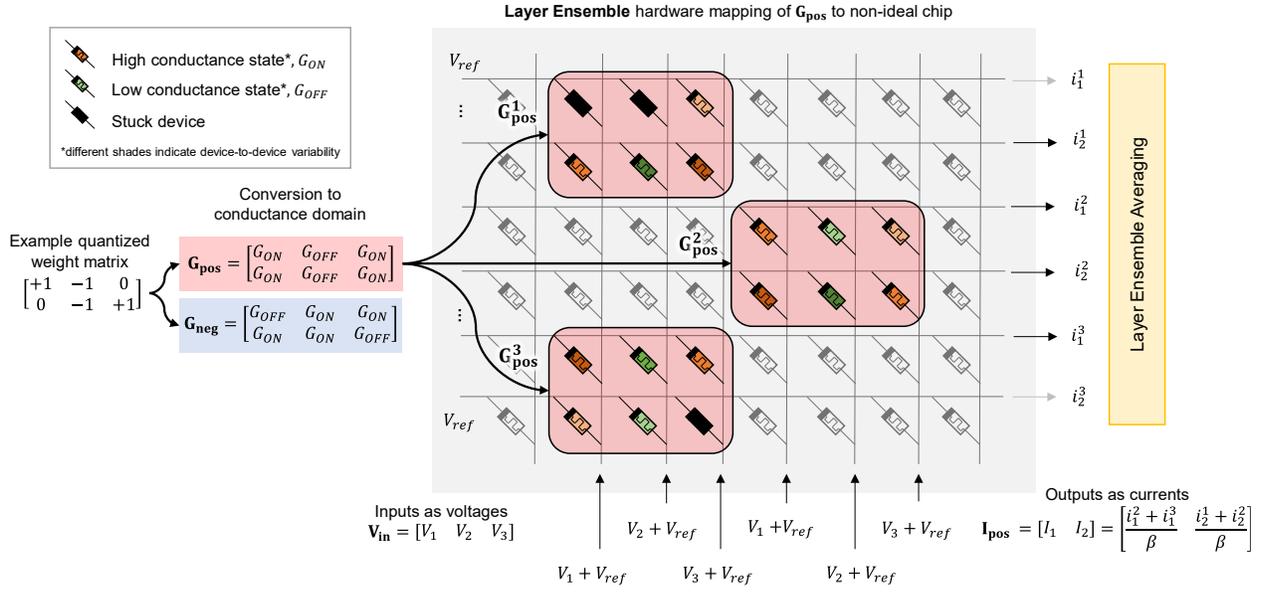

**Figure 2. Layer Ensemble Averaging**. Demonstration of the proposed layer ensemble averaging technique for mapping an example weight matrix to a crossbar of non-ideal memristive devices. The weight matrix is converted into conductance matrices $G_{pos}$ and $G_{neg}$ according to a differential encoding scheme, and each conductance matrix is mapped to a contiguous block of devices on the non-ideal chip according to the layer ensemble mapping algorithm (presented in **Algorithm S1**). The green and orange coloring represents devices in low and high conductance states respectively, with varying shades indicating device-to-device variability, while the black coloring represents stuck or faulty devices. Output currents that contribute to the final averaging process are controlled by the non-defective output parameter of the layer ensemble mapping algorithm, $\beta$. For the presented example, $\beta = 2$. For simplicity, only the layer ensemble mapping for $G_{pos}$ is shown, but both $G_{pos}$ and $G_{neg}$ must be mapped for each layer to fully implement the neural network. While input voltages are applied on columns and output currents are measured on rows of the crossbar for this example, this is not a requirement and can be altered as required.

The algorithm utilized for finding ensemble mappings of a particular weight matrix on the chip is presented in the Supplementary Material as **Algorithm S1**. It is a greedy algorithm requiring conductance map measurements (corresponding to the states to be used for the layer mapping), dimensions of the layer to be mapped, existing mappings to avoid, and an integer value of $\beta$ signifying the number of defect-free copies of rows that the algorithm should find. It proceeds by iteratively adding non-conflicting candidate mappings to its solution until either the solution satisfies the $\beta$ non-defective outputs criteria for each output, or if it fails to find a non-conflicting mapping that would improve the current ensemble mapping solution. The output is a list of ensemble mappings as well as a state vector indicating which, if any, outputs in the ensemble are impacted by a defective device. This information is utilized during the inference operation to mask out currents from defective rows in the ensemble for the layer ensemble averaging process detailed above.

**Experimental Setup**

To experimentally evaluate the effectiveness of layer ensemble averaging, we utilize a custom mixed-signal prototyping system named Daffodil[34]. The full experimental setup is shown in **Figure 3**. It consists of a custom complementary metal-oxide-semiconductor (CMOS)/memristor integrated circuit or chip, a custom mixed-signal PCB named the Daffodil board, and a Zynq-based FPGA development board that acts as the host. The chip with foundry 180 nm CMOS and in-house integrated ReRAM devices is packaged and connects to the PCB via a 21×21 pin Ceramic Pin Grid Array (CPGA) package, and the PCB connects to the FPGA via a fully populated FPGA mezzanine card (FMC) connector.

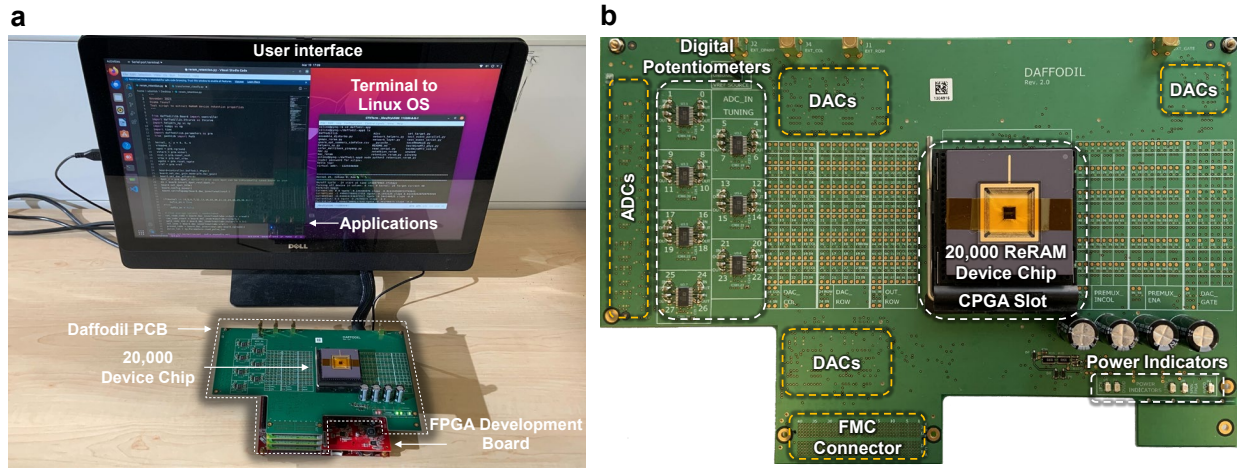

**Figure 3. Overview of the experimental setup.** a) Image of the complete working setup. The FPGA development board interfaces with the Daffodil printed circuit board housing the 20,000 device ReRAM chip (outlined in white) via a FPGA mezzanine card (FMC) connector. The development board also connects to a host system over a serial terminal through which high-level applications can be run. b) A zoomed-in view of the outlined mixed-signal Daffodil board showing major hardware components present on-board. ADCs, DACs, and the FMC connector are located under the board.

At its core, the system is an integrated mixed-signal vector-matrix multiplication processor acting as a neural network accelerator based on emerging two-terminal memristor devices. The crossbar array consists of a total of 20,000 2T-1R devices which are organized into 32 25×25 subarrays, called kernels. Daffodil provides access to the kernels via five external CMOS logic signals. Each kernel is controlled by 75 distinct digital-to-analog converter (DAC) channels, allowing parallel biasing for each kernel row, column, and gate. The PCB also has 25 re-configurable analog-to-digital converter (ADC) channels for measuring currents in parallel on kernels via dedicated transimpedance amplifiers. The reference voltages to these amplifiers can be tuned via a separate DAC channel, and the feedback resistance for each amplifier can be tuned via dedicated digital potentiometers. There are three possible configurations based on ADC and DAC connections suited for different applications, summarized in **Figure S1**. A hard processor on the FPGA development board runs a custom Linux operating system built using Xilinx's PYNQ framework[42]. The programmable logic contains custom register-transfer level code for important use cases such as timed pulse generation, read and write operations, or kernel selection.

The primary library, *daffodil-lib*, acts as the system's control software and design verification framework and handles hardware communication as well as hardware-accurate simulation and modeling. The secondary library, *daffodil-app*, utilizes primitives from *daffodil-lib* and provides complex applications to the user. All experimental results reported in this work were extracted using these libraries. Hardware vector-matrix multiplication operations were implemented on physical crossbars on the 20,000-device chip, and other operations such as network activation functions and layer ensemble averaging were implemented in software directly on the host FPGA development board. The system simplifies studying hardware-aware algorithms for device resistive state tuning and/or neural network mapping, and hardware-software results can be easily verified jointly under the same platform by toggling between the hardware and simulation classes given by *daffodil-lib*. Further details of the prototyping system are presented in supplementary material. An overview of the hardware-software architecture is also presented in **Figure S2**.

**Results**

**Device and Array Characterization**

We summarize the ReRAM device and crossbar array characterization results gathered using the mixed-signal prototyping platform. **Figure 4a** shows a micrograph of the 20,000-device chip. The first step for characterization involves observing current *vs.* voltage (*I-V*) characteristics of these devices. The objective is to identify a set of conductance states to which a maximal percentage of devices on the chip can be written. For our devices, the optimal set was found to comprise of four conductance states, [133, 167, 200, 233] μS, indicating that our devices can be reliably tuned to 2-bit precision. For neural network experiments, the two extreme states from this optimal conductance set, 133 μS and 233 μS, were utilized as $G_{OFF}$ and $G_{ON}$ to maximize the separation for the ternary neural network weights and minimize state overlap, which is a key requirement for implementing a neural network based on these devices.

**Figure 4b** shows retention properties with the four identified conductance states for a single representative device from the chip. The process involves initial tuning or programming of a device to a requested conductance state followed by successive measurements over a specified timeframe. An iterative margin-based programming scheme was utilized where a device is considered successfully programmed if its conductance is read back as $G_{req} \pm \theta$, where $G_{req}$ is the target or requested conductance and $\theta$ is a user-specified margin. A smaller $\theta$ can lead to more precise tuning at the expense of increased difficulty to program as the resolution limitations and noise present in the system (primarily arising from the ADCs and DACs) restrict very precise measurements. A smaller $\theta$ requires more iterations for programming and thus bears a higher chance of wearing out devices. For this reason, a relaxed margin of $\theta = 16.66$ μS was used. The retention data was gathered by tuning the device to a specified state via this scheme, and then reading back its conductance once per second over a period of 2 hours. As can be seen, the devices exhibit relatively stable states, which is a crucial requirement for any neural network. We note here that we found numerous devices that could be tuned to a much larger set of conductance states using our margin-based programming scheme. An example of one such device tuned to a total of 10 unique conductance states is presented in **Figure S3** as part of Supplementary Material. **Figure**

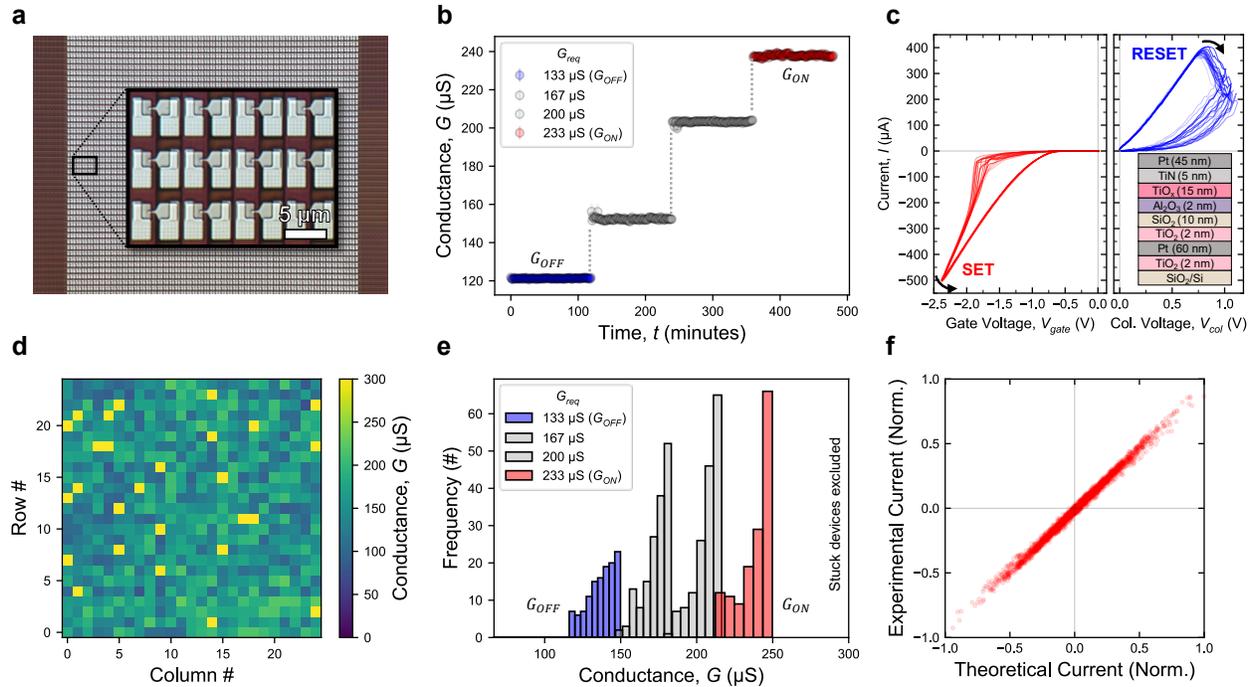

**Figure 4. ReRAM device and array characterization.** a) Zoomed-in device micrograph of a small subarray within the 20,000-device chip. b) Device retention data showcasing effective 2-bit tuning of a representative device. The error bars indicate one standard deviation. States $G_{ON}$ and $G_{OFF}$ are utilized for the neural network implementation. c) Current *vs.* voltage (*I-V*) sweep curves over 20 cycles of a single representative device from the chip showing low cycle-to-cycle variability. The device material stack is also shown. d) Conductance map and e) corresponding conductance distributions of devices in a kernel randomly programmed to one of four conductance states from **Figure 4b**. f) Comparison of theoretical accumulated currents and experimental accumulated currents on the kernel for 100 randomly generated input voltage vectors repeated for a total of 20 iterations showing effective multiply-and-accumulate and vector-matrix multiplication operations. All experimental results are gathered using the Daffodil mixed-signal prototyping platform.

**4c** shows *I-V* sweep curves for a single representative device. The data was gathered over a total of 20 cycles to capture the underlying cycle-to-cycle variability. The SET and RESET phases of the device and the device material stack are highlighted. For the SET phase, we fixed the column voltage at 2.0 V and swept the gate bias. This proved more reliable with our source-drain configuration for the SET operation, with the source bias floating between the memristor and ground. For RESET, we used a traditional column-row voltage sweep with a gate voltage of 3.3 V. The obtained curves show that the devices exhibit acceptable cycle-to-cycle variability. The dynamic range and precision of the current measurements were tuned directly within high-level application code to maximize the effective signal-to-noise ratio.

The specified conductance states were utilized to realize a general vector-matrix multiplication operation on the crossbar where the matrix entries take on one of four values. A single kernel represents the highest degree of physics-based parallelism attainable for our system. For a given kernel, the algorithm for our vector-matrix multiplication experiment proceeds as follows: first, each device is randomly assigned to one of the four conductance states and is tuned to that using our margin-based tuning algorithm. Secondly, a kernel-level map of device conductances is extracted. This information is utilized to directly calculate the theoretical accumulated current for a given vector of input voltages. Finally, a sequence of input voltage vectors (with 25 dimensions corresponding to inputs for the kernel) is generated, with each voltage in each vector selected randomly from the interval $[-0.3, 0.3]$ V (corresponding to $[-V_{read}, V_{read}]$). This random vector of voltages is then applied onto the kernel columns as inputs and the accumulated experimental current outputs are measured on the rows in parallel. These currents are compared with corresponding theoretical outputs. **Figure 4d** shows the conductance map extracted for a representative kernel after the write step of the vector-matrix multiplication operation, and **Figure 4e** shows corresponding device conductances as a histogram. After programming, the devices have two kinds of non-idealities. The first non-ideality arises from the programming scheme. Since the programming scheme is based on a margin, two devices tuned successfully to the same state still have variability. As a result, there is a small degree of overlap between device conductances for neighboring states. The second non-ideality arises from unavoidable physical characteristics of the chip where devices can either be stuck in a high or low conductance state altogether or be electrically shorted. The tuning algorithm fails for these devices. For the results in **Figure 4d**, the tuning succeeds for ~ 96 % of devices showcasing high yield, and only 27 problematic devices were found, all of which were stuck in a high conductance state (bright yellow). **Figure 4f** compares the experimentally measured accumulated currents and the theoretical or expected accumulated currents from this kernel for the vector-matrix multiplication operation. A total of 100 random input vectors were generated and applied, and each accumulated output was measured 20 times to capture read noise and approximate the true mean. The experimental measurements agree very well with theoretical estimates, demonstrating that the vector-matrix multiplication operation is implemented successfully within crossbars in our system. The limited precision of the ADCs and DACs on the prototyping system contribute to the underlying noise observable in the accumulated currents, but the system allows trading off between the dynamic range and precision of the induced currents via a simple resistance tuning process summarized in the Supplementary Material. While this agreement between theoretical and experimental accumulated currents from read-back device states showcases correctness and fidelity of memristive arrays in our system, it does not capture the nuanced disagreements and deviations arising from device defects or variabilities possibly tied to the limited precision and the employed margin-based programming scheme. These disagreements are directly related to the performance gaps between memristive neural networks and their software counterparts. We developed the layer ensemble averaging framework to address this critical issue.

**Exploration of Layer Ensemble Averaging in Simulation**

To gauge the effectiveness of layer ensemble averaging in improving the performance of non-ideal hardware neural networks on the continual learning problem, a simulation testbench was designed where the 20,000-device chip is modelled. **Figure 5a** presents representative kernels from three simulated defective chips exhibiting different levels of device defects. Working devices are modelled as perfectly tunable, while stuck devices are modelled as not being tunable. Since there is no variability among working devices tuned to the same conductance state for a given simulated defective chip, a value of 1 for the non-defective row parameter $\beta$ is sufficient for the ensemble mapping, corresponding to no averaging. Consequently, this testbench directly measures the effectiveness of the layer ensemble averaging methodology for stuck or faulty devices with no device-to-device variations. Given a percentage of devices to model as being stuck, the testbench constructs a simulated defective chip with random devices chosen uniformly modelled as being stuck in the high conductance state until the desired percentage of stuck devices is achieved. Then, the layer ensemble algorithm is executed to generate mappings for the neural network layers. A pre-trained and quantized solution to the continual learning problem is then written to the simulated chip using corresponding layer ensemble mappings, and then the inference operation is carried out on this simulated crossbar using *daffodil-sim*, a simulation framework that emulates the hardware platform end-to-end[34] (further detailed in Supplementary Material).

Since random defects are drawn randomly from a uniform distribution, the layer ensemble mapping and inference operations were repeated for a total of 20 iterations for each percentage of defective devices. **Figures 5b, c, d** summarize neural network inference results on the continual learning problem under investigation for this simulation testbench. **Figure 5b** presents the ensemble size for neural network layers and **Figure 5c** presents the corresponding total count of devices used in the layer ensemble mapping as a function of stuck devices per kernel. On average, as the percentage of stuck devices increases, the ensemble size grows proportional to the dimensions of network layers. This makes sense because it becomes harder to find large contiguous defect-free blocks within kernels when the defects are numerous. As a result, the ensemble size and the total devices required for mapping increase. For the 20,000-device chip and neural network architecture, the layer ensemble mapping operation is successful for up to 35 % stuck devices per kernel. At 40 %, there are instances where the mapping operation fails for the second layer with largest dimensions. In such case, at least one output of the vector-matrix multiplication operation for the second layer would be influenced by a stuck device, resulting in a drop in the network's test accuracy on the continual learning problem. This is evident in **Figure 5d**, which presents the test accuracy of the network averaged for the two tasks. The network can attain software-equivalent inference performance when the layer ensemble mapping operation is successful. When this operation fails, i.e. when at least one output of at least one layer in the ensemble mapping is not guaranteed to be defect-free after the mapping operation, the inference performance of the network degrades severely, dropping well below the performance of a linear solver.

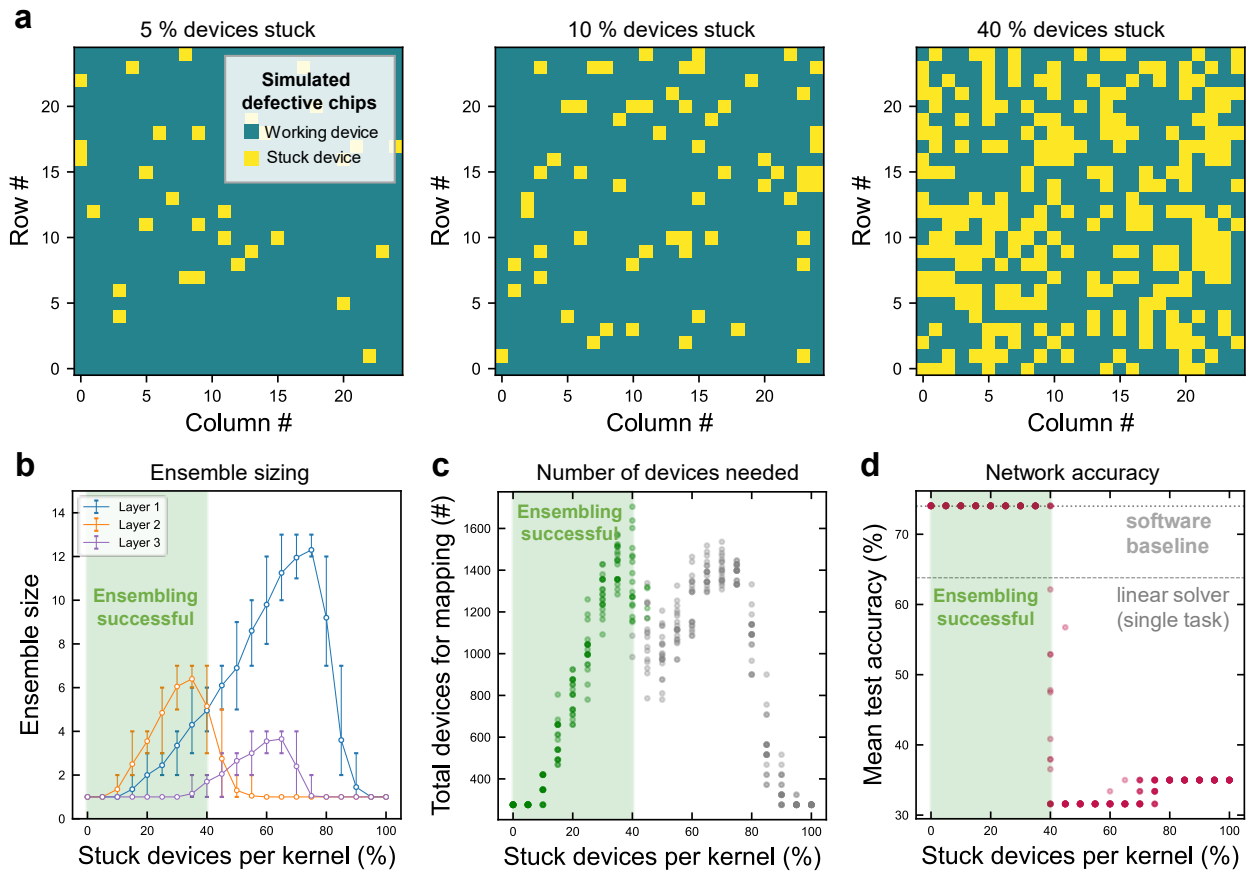

**Figure 5. Simulated investigation of layer ensemble averaging for the continual learning problem.** a) Simulated defective chip kernels with varying levels of faulty or stuck devices with no tunability. The working devices are modelled as perfectly tunable, thus only requiring $\beta = 1$ for the ensemble mapping. b) Average ensemble size for the memristive hardware network on a per-layer basis, c) the total number of devices needed to implement the layer ensembles, and d) testing accuracy on the continual learning problem averaged over the multiple tasks as a function of stuck devices per kernel on simulated defective chips with different defect levels. Successful ensemble regions are highlighted to showcase that the network can reliably attain software-equivalent performance via layer ensemble averaging when ensembling is successful.

These results show that the proposed layer ensemble averaging approach is an effective way to mitigate the impact of faulty memristive devices within the model of our prototyping system. It is likely that different memristive accelerators and systems (i.e., with different kernel sizes, layer dimensions, ADC/DAC precisions, etc.) would exhibit differences in the scaling of ensemble sizes and inference accuracies. However, the effectiveness of layer ensemble averaging would generalize well since it is free from assumptions specific to the architecture of our Daffodil prototyping system. For the given 20,000-device chip and neural network, up to 35 % defects can be reliably tolerated. Thus, the layer ensemble averaging approach gives a way to bring the inference performance of a memristive neural network on par with its software counterpart at the expense of increased devices for mapping.

**Hardware Demonstration of Layer Ensemble Averaging**

Hardware continual learning results with layer ensembles gathered using the mixed-signal prototyping platform are presented in **Figure 6**. We explored different values of the number of non-defective copies of rows parameter ($\beta$). As can be seen in **Figure 6a**, when there is no averaging in the ensemble (i.e., $\beta = 1$) and output currents from only a single non-defective row are used, the inference performance of the network is worse than even a linear solver. However, as $\beta$ increases, the performance increases, approaching the software baseline at $\beta = 3$ (which means that for each output of each layer, 3 non-defective rows are considered for averaging). The benefit gained by increasing $\beta$ further is negligible in terms of inference accuracy. These results show that averaging more currents per output can indeed mitigate the impact of device variabilities and assist the network in performing better. However, improvements in performance diminish as the hardware ensemble network approaches near-software performance, and thus one must decide upon a suitable trade-off since a higher $\beta$ is directly proportional to the total memristive devices required for layer mapping. For reference, results from a non-ensemble ideal model are also presented. These were generated by an end-to-end simulation of our prototyping system with perfectly tunable defect-free devices. **Figure 6b** shows conductance maps of a few kernels involved in the hardware layer ensemble solution mapping for the case where $\beta = 3$. The bounding boxes indicate layer numbers, and the coloring indicates whether the mapped layer is for the positive half of the original weight matrix $\mathbf{G_{pos}}$ (colored red) or negative half $\mathbf{G_{neg}}$ (colored blue). **Figures 6c, d** present the inference performance of the network as a function of $G_{norm}$ for the ideal simulation as well as the hardware layer ensemble network with $\beta = 3$. $G_{norm}$ is the scaling factor for the difference of the average low conductance state from the average high conductance state of the devices participating in the layer mappings. The software accuracy, which serves as the benchmark for our hardware network, and the single-task linear solver accuracy, are also included for reference. As mentioned, the ideal simulation represents perfectly tunable devices with no non-idealities or faults and thus does not require layer ensembles. By comparison, the hardware layer ensemble network includes device non-idealities, as evident from **Figure 6b**.

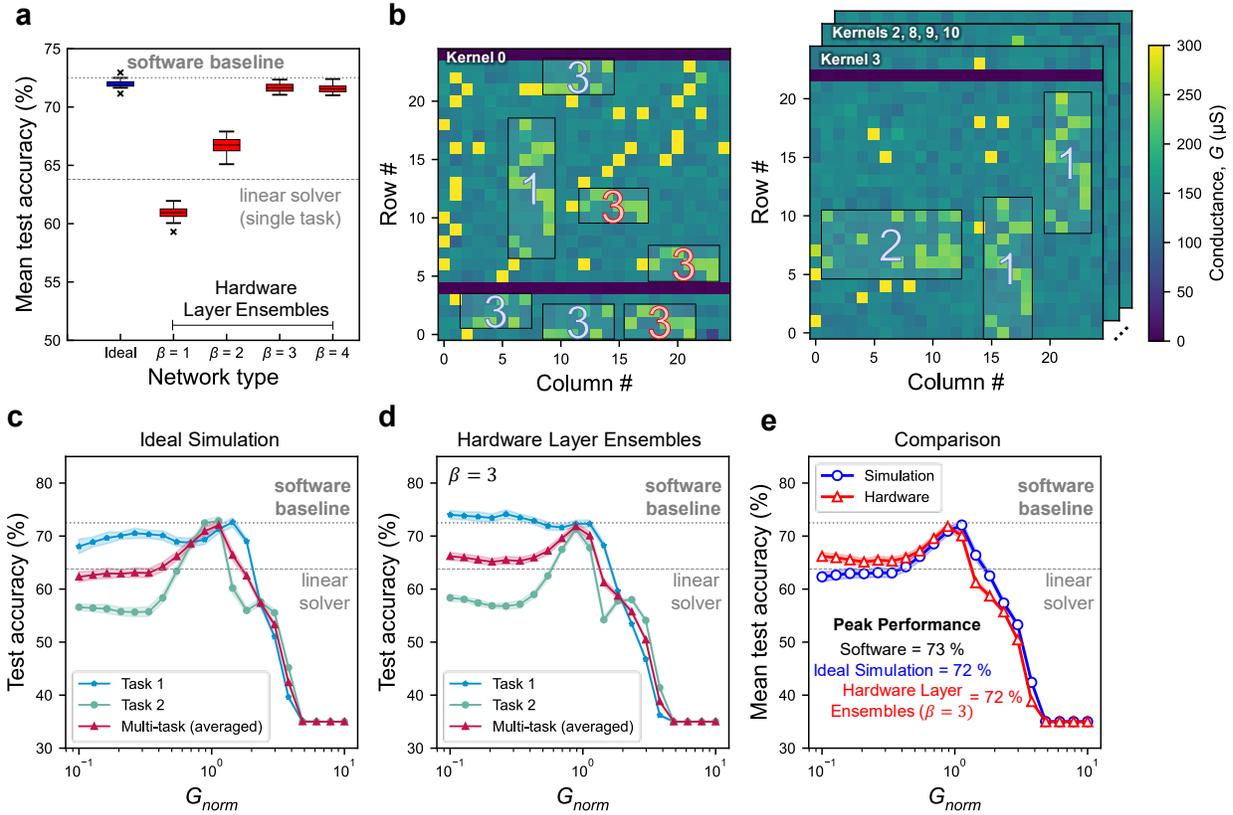

**Figure 6. Hardware demonstration of layer ensemble averaging for the continual learning problem**. a) Average multi-task test set accuracies for hardware neural networks implemented with layer ensemble averaging with increasing values of $\beta$ with $G_{norm} = 1$. Results from an ideal crossbar simulation model (with no layer ensembles and perfect devices) are also included for reference. All inference operations were repeated for a total of 20 iterations to sufficiently capture hardware noise. b) Conductance maps of two kernels from the 20,000-device chip used in the mapping of the solution via hardware layer ensembles with $\beta = 3$. Layer mappings are highlighted by numbered bounding boxes indicating layer number. The red and blue numbering corresponds to mappings of the $\mathbf{G_{pos}}$ and $\mathbf{G_{neg}}$ weight matrices respectively. Test set accuracies as a function of $G_{norm}$ for c) the ideal simulation model and the d) hardware memristive network with layer ensemble averaging using $\beta = 3$. e) Average multi-task test set accuracies from c) and d) together, highlighting that the layer ensemble network can attain near-software and near-ideal simulation performance. Full hardware solution mapping is presented in **Figure S4**. All experimental results are gathered using the Daffodil mixed-signal prototyping platform.

For an ideal memristor-based artificial neural network under the employed mapping scheme, $G_{norm} = 1$ corresponds to the scaling that produces output currents equivalent to software vector-matrix multiplication outputs. For a sufficiently difficult task, the inference performance of a network would be highest at this optimal value. If the output currents of network layers are not optimally scaled, the hardware network would map a different solution than intended, and thus the highest accuracy would likely occur at a point other than $G_{norm} = 1$. As is evident in **Figures 6c and 6d**, the optimal $G_{norm}$ is 1 for the ideal simulation as well as the hardware network with layer ensemble averaging. For the ideal network, this is a direct consequence of the mapping scheme. However, for the hardware network, this shows that layer ensemble averaging effectively corrects vector-matrix multiplication outputs (relative to software) despite being comprised of defective devices, which in turn improves the performance of the memristor-based network up to the software baseline. It can also be seen that there is a considerably large window around $G_{norm} = 1$ where the hardware network can perform better than a linear solver on both tasks. This is important as it highlights that the hardware network can tolerate minor relaxation or device drift defects and attain good performance on the task under investigation without any hyperparameter re-optimizations. **Figure 6e** presents the average multi-task accuracies of the ideal and hardware layer ensemble network together. On average, the hardware layer ensemble network is very close to the performance of the ideal network and can attain near-software inference performance, only deviating 1 % from the software baseline at the optimal $G_{norm}$. All inference experiments were repeated for a total of 20 iterations to capture the impact of hardware noise on the network performance. These results showcase the effectiveness of the proposed layer ensemble averaging methodology. They signify that the $\beta$ parameter can be used to trade-off the number of devices in the solution mapping for network performance improvements in complex multi-task continual learning settings, allowing the network to perform as good as the software baseline regardless of device non-idealities and faults.

**Discussion**

In summary, a 3-layer perceptron network implemented using a defective memristive crossbar capable of attaining near-software (deviating only < 1 % from software) inference performance on a continual learning task has been experimentally demonstrated for the first time. The network was deployed on a 20,000-device chip controlled by a mixed-signal prototyping system designed exclusively for benchmarking two-terminal emerging memory devices for neuromorphic computing applications. Results indicate that layer ensemble averaging can be used to implement a trade-off between inference accuracy and the number of devices used for network layer mappings. Provided enough devices are available for the ensemble mapping to succeed, this trade-off can bring the performance of the memristive network on-par with its software counterpart.

Although the presented continual learning task and the deployed neural network architecture is small, it provides a strong proof-of-concept demonstration of the possibility of performing accurate vector-matrix multiplications in defective memristive crossbars. While this work focused exclusively on utilizing layer ensemble averaging for neural network inference, the methodology is broadly applicable to a variety of domains requiring accurate multiply-and-accumulate and vector-matrix multiplication operations. Dedicated memristive hardware accelerators for such domains could be envisioned in the future owing to their power consumption and energy benefits compared to traditional computing systems. In future work, we will explore continual learning on memristive hardware beyond inference applications. This exploration will involve addressing unique challenges related to memory updates, synaptic plasticity, and preserving previously acquired knowledge. Additionally, we plan to experimentally demonstrate layer ensemble averaging on a large language modeling problem with transformer networks using a more complex memristive crossbar accelerator.

**Acknowledgements**

The authors acknowledge Pi-Feng Chiu for help with the initial FPGA design and Patrick Braganca and Tom Boone for useful advice on nanofabrication and relevant applications.

This work was supported in part by NIST under grant **70NANB22H018**, by Western Digital under grant **ECNS21932N**, by **NSF CAREER** under grant **2239951**, by **AFOSR YIP** under grant **FA9550-23-1-0173**, and by the GW Cross-Disciplinary Research Fund. The authors acknowledge the use of nanofabrication facilities, high-performance computing clusters and advanced IT support from the research technology services at GW and NIST. The NIST authors acknowledge support from the NIST Hardware for AI Program.

**Author Contributions**

O.Y. and G.C.A. conceived the scientific concept. O.Y. developed the core layer ensemble averaging algorithm, and G.C.A., M.W.D. and A.D. helped refine it. O.Y. performed array characterization, neural network training and hardware inference experiments. A.M. designed the integrated circuit. B.D.H. integrated the ReRAM devices. B.D.H. and M.L.B. designed the initial Daffodil board and its API, and K.R. and O.Y. assisted with a subsequent re-design used for this work. K.R. and B.D.H. developed the device tuning algorithm, with help from W.A.B. K.R. and M.F. developed interfacing drivers for the FPGA and daughterboard. M.F., O.Y., and K.R. designed hardware description language code for the FPGA and worked on experimental testing and debugging. O.Y. developed the custom Linux-based operating system for the FPGA development board. All authors assisted in data analysis. O.Y. wrote the initial manuscript and all authors participated in co-editing. G.C.A. and J.J.M. supervised the project.